\documentclass[12pt,epsf]{article}
\usepackage{graphicx}
\usepackage{amsmath,amssymb}
\begin{document}

\def\a{\alpha}
\def\b{\beta}
\def\c{\varepsilon}
\def\d{\delta}
\def\e{\epsilon}
\def\f{\phi}
\def\g{\gamma}
\def\h{\theta}
\def\k{\kappa}
\def\l{\lambda}
\def\m{\mu}
\def\n{\nu}
\def\p{\psi}
\def\q{\partial}
\def\r{\rho}
\def\s{\sigma}
\def\t{\tau}
\def\u{\upsilon}
\def\v{\varphi}
\def\w{\omega}
\def\x{\xi}
\def\y{\eta}
\def\z{\zeta}
\def\D{\Delta}
\def\G{\Gamma}
\def\H{\Theta}
\def\L{\Lambda}
\def\F{\Phi}
\def\P{\Psi}
\def\S{\Sigma}

\def\o{\over}
\def\beq{\begin{eqnarray}}
\def\eeq{\end{eqnarray}}
\newcommand{\gsim}{ \mathop{}_{\textstyle \sim}^{\textstyle >} }
\newcommand{\lsim}{ \mathop{}_{\textstyle \sim}^{\textstyle <} }
\newcommand{\vev}[1]{ \left\langle {#1} \right\rangle }
\newcommand{\bra}[1]{ \langle {#1} | }
\newcommand{\ket}[1]{ | {#1} \rangle }
\newcommand{\EV}{ {\rm eV} }
\newcommand{\KEV}{ {\rm keV} }
\newcommand{\MEV}{ {\rm MeV} }
\newcommand{\GEV}{ {\rm GeV} }
\newcommand{\TEV}{ {\rm TeV} }
\newcommand{\nn}{\nonumber \\}
\def\diag{\mathop{\rm diag}\nolimits}
\def\Spin{\mathop{\rm Spin}}
\def\SO{\mathop{\rm SO}}
\def\O{\mathop{\rm O}}
\def\SU{\mathop{\rm SU}}
\def\U{\mathop{\rm U}}
\def\Sp{\mathop{\rm Sp}}
\def\SL{\mathop{\rm SL}}
\def\tr{\mathop{\rm tr}}

\def\IJMP{Int.~J.~Mod.~Phys. }
\def\MPL{Mod.~Phys.~Lett. }
\def\NP{Nucl.~Phys. }
\def\PL{Phys.~Lett. }
\def\PR{Phys.~Rev. }
\def\PRL{Phys.~Rev.~Lett. }
\def\PTP{Prog.~Theor.~Phys. }
\def\ZP{Z.~Phys. }


\baselineskip 0.7cm

\begin{titlepage}

\begin{flushright}
IPMU14-0091
\end{flushright}

\vskip 1.35cm
\begin{center}
{\large \bf Upper Bounds on Gluino, Squark and Higgisino Masses \\
in the Focus Point Gaugino Mediation \\
with a Mild Fine Tuning $\Delta \le 100$
}
\vskip 1.2cm
{\bf   Tsutomu T. Yanagida and Norimi Yokozaki}
\vskip 0.4cm

{\it  Kavli IPMU (WPI), University of Tokyo, Kashiwa, Chiba 277-8583, Japan }

\vskip 1.5cm

\abstract{We show that upper bounds on the masses for gluino, squarks and higgsino are
$m_{\rm gluino} \le \,5.5{\rm TeV}, m_{\rm squark} \le 4.7\,{\rm TeV}$ and $m_{\rm higgsino} \le 650\,{\rm GeV}$
in a focus point gaugino mediation. Here, we impose a mild fine tuning $\Delta \le 100$.
This result shows that it is very challenging for the LHC to exclude the focus point gaugino mediation with the mild fine tuning. However, the ILC may have a potential for excluding the focus point gaugino mediation with such a mild fine tuning. 
It is also shown that vector-like matters reduce the required masses of the squark (stop) and gluino to explain the observed Higgs boson mass and enhance the testability of the model at the LHC. The fine-tuning is still kept mild.
%
}

\end{center}
\end{titlepage}

\setcounter{page}{2}

\section{Introduction}
Gaugino dominated supersymmetry (SUSY) breaking mediation, so-called {\it gaugino mediation}, had been proposed as a solution to the flavor-changing neutral current (FCNC) problem~\cite{IKYY, gaugino_ext}, long time ago. In the gaugino mediation model, masses of squarks and sleptons are assumed to be suppressed compared to the gaugino masses at the high energy scale. 
The scalar masses at the weak scale are generated from gaugino loop contributions.  
Since the gaugino contributions to masses of squarks and sleptons are always flavor independent, SUSY contributions to FCNC processes such as $K$ meson mixing and $\mu \to e \gamma$ are suppressed.\footnote{For theses SUSY contributions, see Refs.~\cite{kkbar} ($K$ meson mixing) and  \cite{lfv1, lfv2} ($\mu \to e \gamma$).} These flavor changing processes are serious obstacles to the low-energy SUSY.

If the SUSY is a solution to the hierarchy problem, focus point scenarios~\cite{focus_org, focus, martin_gaugino}
are now attractive. This is because the relatively heavy Higgs boson of around 125 GeV suggests that SUSY particles are heavier than a few TeV~\cite{OYY}, along with the non-observations of the SUSY particles at the LHC.
In the focus point scenarios, the EWSB can be explained naturally 
even if the SUSY particles are much heavier than the EWSB scale.
It had been known in the general framework of gravity mediation that gaugino contributions to the Higgs potential have a focus point behavior at the electroweak scale if gaugino masses are non universal at the GUT scale \cite{focus}.

Motivated by those considerations above, we proposed, recently, a gaugino dominated SUSY breaking scenario with non-universal gaugino masses called as "Focus Point Gaugino Mediation" \cite{yy_focus1, yy_focus2}.\footnote{
See also Ref.~\cite{martin_gaugino}.} We showed that we can obtain the correct electroweak symmetry breaking with a much mild fine tuning even though soft masses of SUSY particles are in a region of a several TeV, thanks to the presence of a focus point. We also show that this focus point gaugino mediation (FPGM) model can easily explain the observed mass of the higgs boson, 
\beq {\rm ATLAS}: \, 125.5\pm 0.2 ^{+0.6} _{-0.6} \  {\rm GeV} ~\cite{ATLAS}, \ \   
{\rm CMS}:  \, 125.7\pm 0.3 \pm 0.3 \  {\rm GeV} ~\cite{CMS} , \nonumber  
\eeq in accord with a mild fine-tuning $\Delta\le 100$ (see Eq.(\ref{eq:def_delta}) for the definition of $\Delta$).

The purpose of this letter is to give upper bounds on SUSY particle masses in the FPGM requiring a mild fine tuning less than 1$\%$ ($\Delta\le 100$) and discuss discovery or exclusion potential of the model at LHC and/or ILC.

\section{Focus point gaugino mediation}

In the focus point gaugino mediation, the EWSB scale becomes relatively insensitive to the gaugino mass parameter at the GUT scale, provided that the ratios of the wino mass $M_2$ to the gluino mass $M_3$ is $M_3/M_2 \sim 8/3$. {The gaugino mass ratio is assumed to be determined by more fundamental physics; non-universal gaugino masses with fixed ratios arise as results of a product group unification model~\cite{yy_focus1}, an anomaly of a discrete R-symmetry~\cite{yy_focus2}, and so on. (See Refs.\cite{PGU} for the details of the product group unification models.)}
The Higgs soft SUSY breaking masses as well as the squark and sleptons masses are generated by the gaugino loops. Therefore the EWSB scale is determined by only the gaugino mass parameters and $\mu$ parameter.
The vacuum expectation values of the up-type Higgs and down-type Higgs and their ratio are determined by following two conditions:
\beq
\frac{m_{\hat Z}^2}{2} \simeq \frac{(m_{H_d}^2 + \frac{1}{2v_d}\frac{\partial\Delta V}{\partial v_d}) - (m_{H_u}^2 +\frac{1}{2v_u}\frac{\partial\Delta V}{\partial v_u})\tan^2\beta }{\tan^2 \beta -1} - \mu^2, \label{eq:ewsb_A}\\
 B\mu (\tan\beta + \cot \beta) \simeq \left(m_{H_u}^2 + \frac{1}{2v_d}\frac{\partial\Delta V}{\partial v_u}+ m_{H_d}^2 +\frac{1}{2v_u}\frac{\partial\Delta V}{\partial v_u}+ 2 \mu^2 \right)  \label{eq:ewsb_B},
\eeq
where $v_u =\left<H_u^0\right>$ ($v_d=\left<H_d^0\right>$) is the vacuum expectation value of the up type (down type) Higgs and $\Delta V$ is the one-loop corrections to the Higgs potential. Here, $\tan\beta=v_u/v_d$. The soft masses of  the up-type and down-type Higgs are denoted by $m_{H_u}$ and $m_{H_d}$, respectively. The Higgsino mass parameter is denoted by $\mu$. The EWSB scale should satisfies the experimental value as $m_{\hat Z} \simeq  91.2$ GeV~\cite{pdg}.

In order to estimate the sensitivity of the EWSB scale with respect to the gaugino mass parameter, we adapt the following fine-tuning measure~\cite{ft_measure}:
\beq
\Delta = {\rm{max}} (|\Delta_a|), \ \Delta_a = \left(\frac{\ln m_{\hat Z}}{\ln \mu_0}, \, \frac{\ln m_{\hat Z}}{\ln M_{1/2}}, \frac{\ln m_{\hat Z}}{\ln B_0} \right) \label{eq:def_delta},
\eeq
where $\mu_0$ and $B_0$ are the Higgsino mass parameter and the Higgs B-parameter at the GUT scale, respectively. We assume that the ratios of the gaugino mass parameters are fixed at the GUT scale.
\beq
M_1/M_2=r_1, \ M_3/M_2=r_3, \ M_2=M_{1/2},
\eeq
where $M_1$ is the bino mass at the GUT scale. Since the focus point behavior is insensitive to $M_1$, we take $r_1=1$ in our numerical calculations.

\begin{figure}[t]
\begin{center}
\includegraphics[scale=1.07]{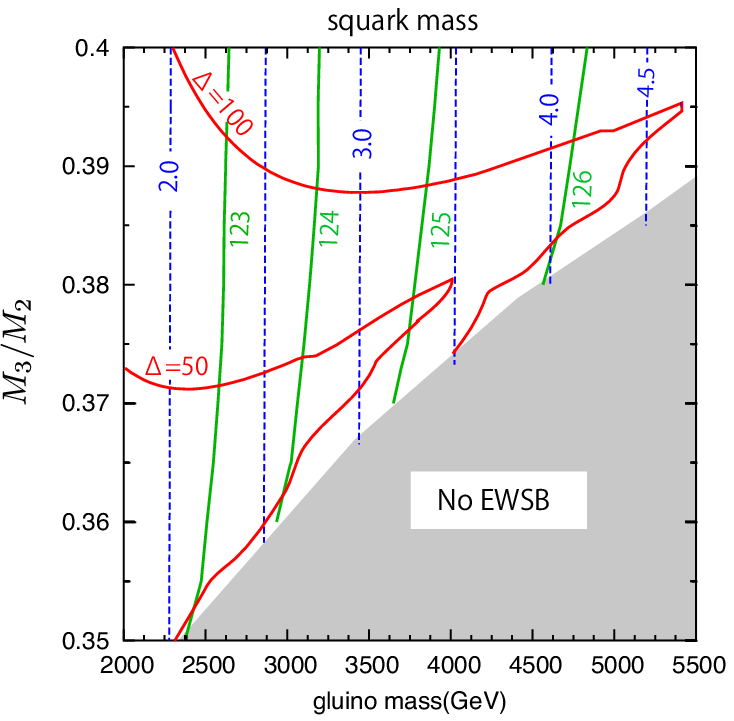}
\includegraphics[scale=1.07]{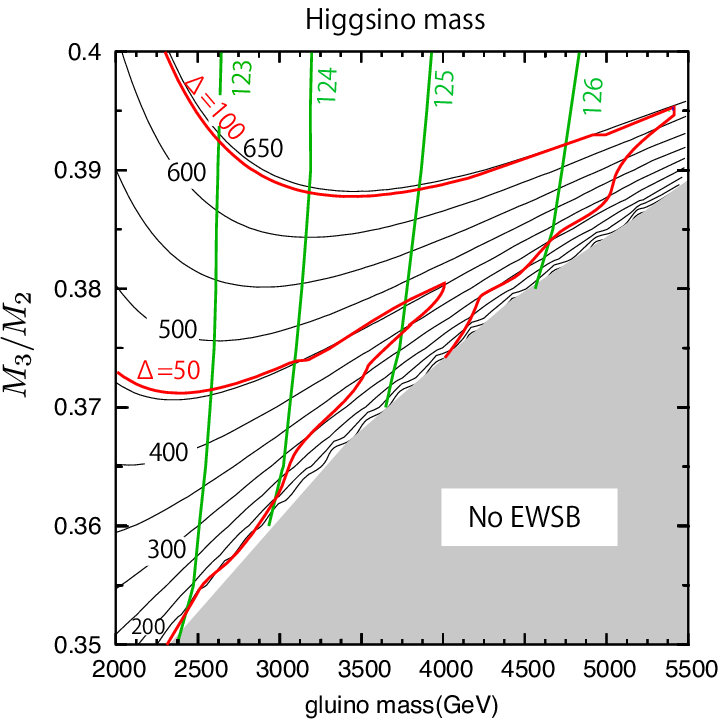}
\caption{The contours of the squark mass (blue dashed lines) and the Higgsino mass (black solid line).
The squark mass (Higgsino mass) is shown in the unit of TeV (GeV).
The red solid line and green solid line correspond to $\Delta$ and $m_h$ (GeV), respectively.
The gray region is excluded because of the unsuccessful electroweak symmetry breaking.
The bino mass at the GUT scale is taken as $M_1=M_2$. Here $\mu<0$, $\tan\beta=20$, $m_t=173.3$ GeV and $\alpha_S(m_Z)=0.1184$.
}
\label{fig:stop}
\end{center}
\end{figure}

In the universal gaugino mass case $r_1=r_3=1$, the Higgs boson mass of $m_h = 125$ GeV is explained with $\Delta \simeq 1300$; the required tuning is more than $0.1$  \% level. However, in the non-universal case, the required fine-tuning is reduced significantly. In Fig.~\ref{fig:stop}, the contours of the Higgs boson mass (green) and $\Delta$ (red) are shown. The Higgs boson mass is calculated using {\tt FeynHiggs\,2.10.0}~\cite{feynhiggs, feynhiggs_higher}, which includes higher order corrections beyond 2-loop level~\cite{feynhiggs_higher}. We use {\tt SuSpect}~\cite{suspect} to evaluate a SUSY mass spectrum and 2-loop renormalization group evolutions. In the focus point gaugino mediation, $m_h \simeq 125$ GeV is explained with $\Delta \sim 50$, when the gaugino mass ratios are set to be $r_3\sim3/8, \ r_1=1$.  Here, we take $\mu<0$, since it can be consistent with $B_0=0$ for $\tan\beta=\mathcal{O}(10)$. Notice that the gaugino mediation model with $B_0=0$ is completely free from the SUSY CP problem. 

\section{LHC and ILC}

The focus point gaugino mediation may be 
difficult to be excluded at the LHC, since the squarks and gluino are too heavy even when the mild fine-tuning  $\Delta=50-100$ is imposed. 
The squark (blue dashed lines) and gluino masses are shown in Fig.~\ref{fig:stop} (left panel). 
The upper bounds on the gluino and squark masses are $m_{\tilde{g}} \lesssim 4.0\, (5.5)$ TeV and $m_{\tilde{q}} \lesssim 3.5\, (4.7)$ TeV for $\Delta < 50\, (100)$, respectively. Here, $m_{\tilde{q}}$ is the  mass of the lightest 1st/2nd generation squark. 

At the 14 TeV LHC, the squark and gluino masses up to $3.2$ TeV and $3.5$ TeV can be covered using 3000 fb$^{-1}$ data~\cite{atlas_3000}. However, the upper bounds on the gluino and squark masses with $\Delta = 50-100$ are beyond the reach of the LHC. Therefore the FPGM model with the mild fine-tuning is hard to be excluded. 
 Moreover, the Higgs boson mass of 125 GeV is explained with $m_{\tilde{g}} \simeq 3.7$ TeV and $m_{\tilde{q}} \simeq 3.2$ TeV. As a result, it is challenging to discover the SUSY particles in the minimal supersymmetry standard model (MSSM).

{
On the other hand, the Higgsino lighter than 690 GeV may be excluded at the ILC, by measuring the cross section $\sigma(e^- e^+ \to \mu^- \mu^+)$ very precisely. As shown in Fig.~\ref{fig:stop}, the Higgsino mass is bounded from above as $\mu < 450\,(650)$ GeV for $\Delta < 50\,(100)$. With this mass of the Higgsino, the gauge couplings change at $\mathcal{O}(0.1\%)$ level as 
\beq
\frac{g_{2}^2(q^2)_{WH}}{g_{2}^2(q^2)} = \left[ 1 + \frac{g_2^2 (q^2)}{4\pi^2} \int_0^1 dx \, x(1-x)  \ln \left(\frac{\mu^2}{\mu^2-x(1-x)q^2} \right) \right]^{-1},
\eeq
where $g_2(q^2)_{WH}$ ($g_2(q^2)$) is the gauge coupling with (without) the Higgsino loop correction (see Appendix A). Taking $\sqrt{q^2}= 500\, {\rm GeV}$ ($1000 \, {\rm GeV})$, the Higgsino with the mass of $\mu \simeq 340$ GeV ($690$ GeV) changes the coupling by 0.1\%.  Similarly, we have
\beq
\frac{g_{1}^2(q^2)_{WH}}{g_{1}^2(q^2)} = \left[ 1 + \frac{3}{5} \frac{g_1^2 (q^2)}{4\pi^2} \int_0^1 dx \, x(1-x)  \ln \left(\frac{\mu^2}{\mu^2-x(1-x)q^2} \right) \right]^{-1},
\eeq
where $g_1(q^2)$ is the GUT normalized $U(1)_Y$ gauge coupling. This gives 0.03\% change in the $U(1)_Y$ gauge coupling at the weak scale. 
Therefore, if the ILC with $\sqrt{s}=1$\, TeV can measure $\sigma(e^- e^+ \to \mu^- \mu^+)$ at $0.1\%$ level using polarized beams, the Higgsino mass up to 690\,GeV can be excluded, even if the Higgsino is not produced directly at the ILC.}

Finally let us comment on the case where vector-like matters are added to the MSSM. The presence of the additional vector-like matters is motivated by, for instance, the existence of the non-anomalous discrete R-symmetry~\cite{znr_nonanomalous}. With the vector-like matters, the gluino and squarks become light compared to those in MSSM.
In this case, the squarks and gluino can be discovered at the LHC.

We introduce $N_5$ pairs of the vector-like matters which are ${\bf 5}$ and $\bar{\bf 5}$ representation of the $SU(5)$ GUT gauge group. The Yukawa couplings between vector-like matters and MSSM matters are assumed to be suppressed. Due to the presence of the vector-like matters, the renormalization group equations (RGEs), especially for gauge couplings and gaugino masses, change (see Appendix B). These changes lead to the significant changes in the SUSY mass spectrum, and the squark (stop) and gluino mass {are reduced} for a given Higgs boson mass~\cite{vector_gaugino}.
The fine-tuning measure is defined with inclusion of the mass of the vector-like multiplets $M_5$. 
\beq
\Delta = {\rm{max}} (|\Delta_a|), \ \Delta_a = \left(\frac{\ln m_{\hat Z}}{\ln \mu^0}, \, \frac{\ln m_{\hat Z}}{\ln M_{1/2}}, \frac{\ln m_{\hat Z}}{\ln B_0}, \, \frac{\ln m_{\hat Z}}{\ln M_5} \right).
\eeq
Note that the sensitivity of $m_{\hat Z}$ with respect to $M_{5}$ is rather weak as $\Delta \lesssim 10$ in the parameter space of interest. 

\begin{figure}[t]
\begin{center}
\includegraphics[scale=1.07]{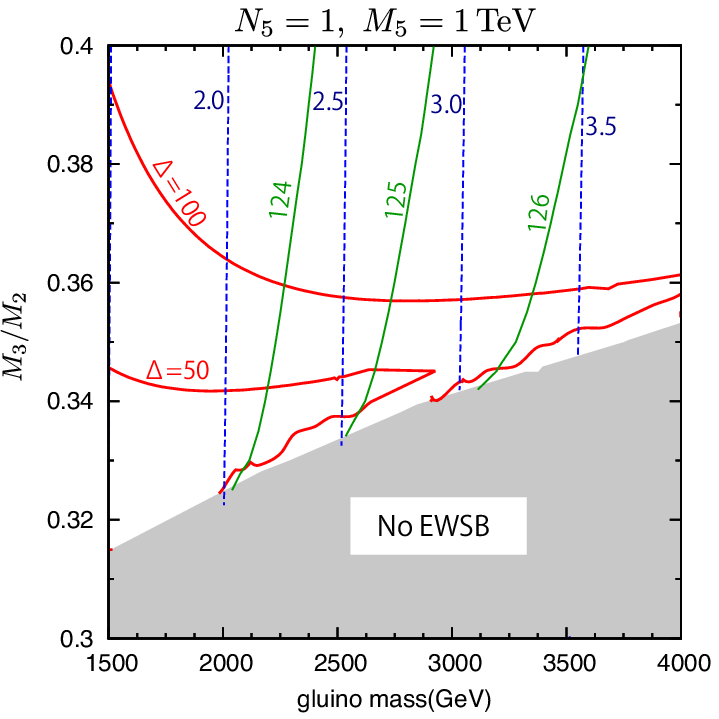}
\includegraphics[scale=1.07]{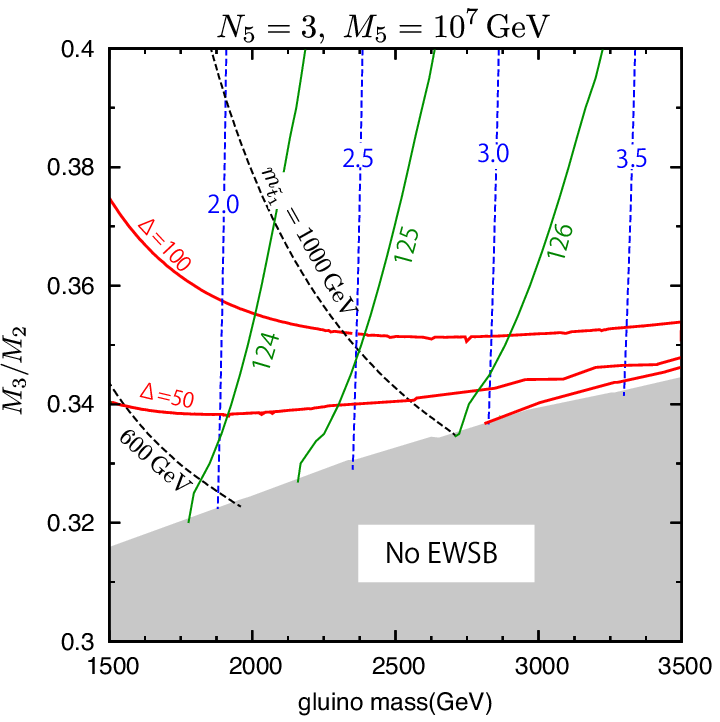}
\caption{The contours of the squark mass (blue) and the Higgs boson mass with vector-like matter(s).
The number and mass of the vector-like multiplets are taken as $N_5=1$ and $M_5=10^3$ GeV ($N_5=3$ and $M_5=10^7$ GeV) in the left (right) panel.
}
\label{fig:extra}
\end{center}
\end{figure}

In Fig.~\ref{fig:extra}, the contours of the squark mass, $\Delta$ and $m_h$ are shown with $N_5$ pairs of vector-like matters included. In both cases ($N_5=1, M_5=1$ TeV and $N_5=3, M_5=10^7$ GeV), $m_h=125$ GeV is explained with a mild fine-tuning $\Delta < 50$. In the first case ($N_5=1, M_5=1$ TeV),  $m_{\tilde{g}} \simeq m_{\tilde{q}} \simeq 2.6$ TeV is consistent with $m_{h} \simeq 125$\,GeV (see left panel). With three pairs of the vector-like matters of $M_5=10^7$ GeV, the observed Higgs boson mass is consistent with $m_{\tilde{g}} \simeq 2.2$ TeV (see right panels). Since the gluino mass $m_{\tilde{g}} \simeq 2.2-2.6$ TeV is within the reach of the LHC, a discovery of the gluino may suggest the presence of the vector-like matters. 
 Moreover, the lightest stop can be light as 800-1000 GeV for $N_5=3$ and $M_5=10^7$ GeV. In this case, the stop can be produced directly at the 14 TeV LHC.

\section{Conclusion and discussion}
We have shown that the upper bounds on the gluino and squark masses are $m_{\tilde{g}}< 5.5$ TeV and $m_{\tilde{q}}< 4.7$ TeV ($m_{\tilde{g}}< 4.0$ TeV and $m_{\tilde{q}}< 3.5$ TeV) in the focus point gaugino mediation model with a mild fine-tuning, $\Delta <100 \,(50)$. These upper bounds show that it is difficult to exclude the FPGM model satisfying a mild fine-tuning at the LHC with $\sqrt{s}=14$ TeV. 

On the other hand, the ILC may have a potential to exclude the FPGM model. A squared of a running gauge coupling changes by $\mathcal{O}(0.1\%)$ level with radiative corrections from the Higgsinos. This change of the gauge couplings reflects a deviation in a cross section $\sigma(e^+ e^- \to \mu^+ \mu^-)$ from the standard model prediction. If the ILC can measure this cross section precisely as 0.1\% level, the Higgsino with the mass less than 650 GeV corresponding to $\Delta<100$ can be excluded.

We have also shown that if the vector-like matters exist at the TeV or at an intermediate scale, the gluino and squark become light as $m_{\tilde{g}} \sim  2.5$ TeV and $m_{\tilde{q}} \sim 2.5$ TeV, and they can be in the region accessible to  the LHC. We find that the fine-tuning is still kept mild even with the presence of those extra-matters.

\section*{Acknowledgment}
We thank Shigeki Matsumoto for useful discussions. 
This work was supported by JSPS KAKENHI Grant 
No.~22244021 (T.T.Y), and also by World Premier International Research Center Initiative
(WPI Initiative), MEXT, Japan. The work of NY is supported in part by JSPS Research Fellowships for Young Scientists.


\appendix
\section{Running gauge coupling}

The existence of the chiral fermion, the running gauge coupling is given by
\beq
\alpha_{\rm eff}(q^2) &=& \frac{\alpha}{1-\hat{\Pi}_2(q^2)}, \\ \nonumber
\hat{\Pi}_2 (q^2) &=& -\frac{3 b \,\alpha}{2\pi} \int_0^1 dx \,x(1-x) \ln \left(\frac{m^2}{m^2-x(1-x)q^2} \right).
\eeq
where $b=(2/3) T(R)$ and $T(R)$ is the Dynkin index of the representation $R$. The mass of the fermion is denoted by $m$. As for the Higgsinos, one-loop corrections give $b=2/5$ and $2/3$ for the GUT normalized $U(1)_Y$ couplings and $SU(2)_L$, respectively.
In the short distance limit $-q^2 \gg m^2$, we have
\beq
\alpha_{\rm eff}(q^2) = \frac{\alpha}{\displaystyle 1-\frac{b\,\alpha}{4\pi} \ln \frac{-q^2}{C m^2}},
\eeq
with $C=\exp(5/3)$.

The ratio of the gauge coupling constants are given by
\beq
\frac{\alpha_{\rm SM+NP} (q^2)}{\alpha_{\rm SM} (q^2)} \simeq \left[1 - \frac{\alpha_{\rm SM}(q^2)}{\alpha}\Pi_{\rm NP} (q^2) \right]^{-1}.
\eeq

%
%
%
%

\section{The renormalization group equations with vector-like matters}
In this appendix, we give two-loop renormalization group equations in $\overline{\rm DR}$ scheme with vector-like multiplets. Here, we define the renormalization scale as $t=\ln Q$. The vector-like matters are introduced as ${\bf 5}=(\bar{L}', {D}')$ and $\bar{\bf 5}=(L', {\bar D}')$ representations in $SU(5)_{\rm GUT}$ gauge group. At the one-loop level,  the renormalization group equations (RGEs) of a model with $N_5$ pairs of the vector-like multiplets change from those in the MSSM:
\beq
\frac{d g_a}{d t} = \frac{b_a+N_5}{16\pi^2} g_i^3, \ \ (b_1, b_2, b_3)=(33/5, 1, -3),
\eeq
\beq
\frac{d M_a}{d t} = \frac{b_a+N_5}{8\pi^2} g_i^2 M_i,
\eeq
\beq
\frac{d m_i^2}{d t} = \left(\frac{d m_i^2}{d t}\right)_{\rm MSSM} + \frac{g_1^2}{8\pi^2} (3/5) Q_i N_5 (m_{\bar{L}'}^2-m_{{L}'}^2 - m_{D'}^2+m_{\bar{D}'}^2),
\eeq
where $Q_{i}$ is a hyper-charge of the chiral matter multiplet. We denote the gauge coupling, gaugino mass and the scalar mass squared as $g_a$, $M_a$ and $m_i^2$.
In gaugino mediation, $(m_{\bar{L}'}^2-m_{{L}'}^2 - m_{D'}^2+m_{\bar{D}'}^2) \simeq 0$.

Following~\cite{martin_rge}, the RGEs of the gauge couplings at the two-loop level are given by
\beq
\frac{d g_a}{dt} =  \frac{B_2^{ab}}{(16\pi^2)^2} {g_a^3} g_b^2,
\eeq
where 
\beq
B_2^{ab}=
\left(
\begin{array}{ccc}
\displaystyle \frac{199}{25} + \frac{7}{15}N_5 & \displaystyle \frac{27}{5}+ \frac{9}{5}N_5 & \displaystyle \frac{88}{5} +\frac{32}{15}N_5 \\ \\
\displaystyle \frac{9}{5} + \frac{3}{5}N_5 & \displaystyle 25 +7N_5 & 24 \\ \\
\displaystyle \frac{11}{5} + \frac{4}{15} N_5 & 9 & \displaystyle 14 + \frac{34}{3}N_5 
\end{array}
\right).
\eeq
With this $B_2^{ab}$, the RGEs of the gaugino masses are written as 
\beq
\frac{d M_a}{d t} = \frac{2g_a^2}{(16\pi^2)^2} B_2^{ab} g_b^2\,(M_a + M_b).
\eeq
The new part of the RGE of the top Yukawa coupling through the change of anomalous dimensions is given by
\beq
\frac{dY_t}{dt} = \frac{Y_t}{(16\pi^2)^2}\left(\frac{13}{15} N_5 g_1^4 + 3 N_5 g_2^4 + \frac{16}{3} N_5 g_3^4\right),
\eeq
and that of the corresponding scalar trilinear coupling is
\beq
\frac{d A_t}{dt} = \frac{(-4)}{(16\pi^2)^2}\left(\frac{13}{15} N_5 g_1^4 M_1 +3 N_5 g_2^4 M_2 +  \frac{16}{3} N_5 g_3^4 M_3 \right).
\eeq

The scalar masses receive negative corrections from the vector-like multiplets. The two-loop renormalization group equations for the scalar masses change as~\cite{martin_rge}
\beq
\frac{d m_{L}^2}{dt} &=& \frac{1}{(16\pi^2)^2}\left[3 g_2^2 \delta \sigma_2 +  \frac{3}{5}g_1^2 \delta \sigma_1 - \frac{6}{5}g_1^2 \delta S' + \frac{18}{5} N_5 g_1^4 M_1^2  + 18 N_5 g_2^4 M_2^2 \right], \nn
\frac{d m_{\bar{E}}^2}{dt} &=& \frac{1}{(16\pi^2)^2}\left[ \frac{12}{5}g_1^2 \delta \sigma_1  + \frac{12}{5}g_1^2 \delta S' + \frac{72}{5} N_5 g_1^4 M_1^2 \right], \nn
\frac{d m_{Q}^2}{dt} &=& \frac{1}{(16\pi^2)^2}\left[\frac{16}{3} g_3^2 \delta \sigma_3 + 3 g_2^2 \delta \sigma_2 +  \frac{1}{15}g_1^2 \delta \sigma_1 + \frac{2}{5} g_1^2 \delta S' \right. \nn
&+& \left. \frac{2}{5} N_5 g_1^4 M_1^2  + 18 N_5 g_2^4 M_2^2 + 32 N_5 g_3^4 M_3^2 \right], \nn
\frac{d m_{\bar{U}}^2}{dt} &=& \frac{1}{(16\pi^2)^2}\left[\frac{16}{3} g_3^2 \delta \sigma_3 +  \frac{16}{15} g_1^2 \delta \sigma_1 - \frac{8}{5}g_1^2\delta S' + \frac{32}{5}N_5 g_1^4 M_1^2  + 32 N_5 g_3^4 M_3^2 \right],\nn
\frac{d m_{\bar{D}}^2}{dt} &=& \frac{1}{(16\pi^2)^2}\left[\frac{16}{3} g_3^2 \delta \sigma_3 +  \frac{4}{15}g_1^2 \delta \sigma_1 + \frac{4}{5}g_1^2 \delta S' + \frac{8}{5} N_5 g_1^4 M_1^2  + 32 N_5 g_3^4 M_3^2 \right],\nn
\frac{d m_{H_u}^2}{dt} &=& \frac{1}{(16\pi^2)^2}\left[3 g_2^2 \delta \sigma_2 +  \frac{3}{5}g_1^2 \delta \sigma_1 + \frac{6}{5}g_1^2 \delta S' + \frac{18}{5} N_5 g_1^4 M_1^2  + 18 N_5 g_2^4 M_2^2 \right],\nn
\frac{d m_{H_d}^2}{dt} &=& \frac{1}{(16\pi^2)^2}\left[3 g_2^2 \delta \sigma_2 +  \frac{3}{5}g_1^2 \delta \sigma_1 - \frac{6}{5}g_1^2 \delta S' + \frac{18}{5} N_5 g_1^4 M_1^2  + 18 N_5 g_2^4 M_2^2 \right],\nn
\eeq
where 
\beq
\delta \sigma_3 &=& g_3^2 N_5 (m_{\bar{D}'}^2+m_{{D}'}^2), \ 
\ \delta \sigma_2 = g_2^2 N_5 ( m_{L'}^2 + m_{\bar{L}'}^2), \ \nn
\delta \sigma_1 &=& (1/5) g_1^2 N_5 ( 3 m_{L'}^2 + 3m_{\bar{L}'}^2 + 2 m_{\bar{D}'}^2 + 2 m_{{D}'}^2), \nn
\delta S' &=& N_5 \left[ \left(\frac{3}{2} g_2^2+\frac{3}{10} g_1^2 \right) ( m_{{\bar L}'}^2 -m_{L'}^2 ) + \left(\frac{8}{3} g_3^2+\frac{2}{15} g_1^2 \right) ( m_{{\bar D}'}^2 -m_{D'}^2 ) \right].
\eeq
Here, we have given only terms which arise from $N_5$ pairs of the vector-like matter multiplets.
In gaugino mediation, $\delta S' \simeq 0$.

The Higgsino mass parameter also receives corrections:
\beq
\frac{d \mu}{dt} &=& \mu \left(-\frac{1}{2} \right) \left(\frac{d \ln Z_{H_u}}{d t} + \frac{d \ln Z_{H_d}}{d t}\right), \nn
&=& \frac{\mu}{(16\pi^2)^2} \left(\frac{3}{2} N_5 g_2^4 + \frac{1}{15} N_5 g_1^4 \right).
\eeq

%
%
%


\begin{thebibliography}{99}

\bibitem{IKYY} 
  K.~Inoue, M.~Kawasaki, M.~Yamaguchi and T.~Yanagida,
  Phys.\ Rev.\ D {\bf 45}, 328 (1992).
  
   \bibitem{gaugino_ext}
  D.~E.~Kaplan, G.~D.~Kribs and M.~Schmaltz,
  Phys.\ Rev.\ D {\bf 62}, 035010 (2000)
  [hep-ph/9911293];
   Z.~Chacko, M.~A.~Luty, A.~E.~Nelson and E.~Ponton,
  JHEP {\bf 0001}, 003 (2000)
  [hep-ph/9911323].

  \bibitem{kkbar}
  F.~Gabbiani, E.~Gabrielli, A.~Masiero and L.~Silvestrini,
  Nucl.\ Phys.\ B {\bf 477}, 321 (1996)
  [hep-ph/9604387].

 \bibitem{lfv1}
   R.~Barbieri and L.~J.~Hall,
  Phys.\ Lett.\ B {\bf 338} (1994) 212
  [hep-ph/9408406];
  R.~Barbieri, L.~J.~Hall and A.~Strumia,
  Nucl.\ Phys.\ B {\bf 445}, 219 (1995)
  [hep-ph/9501334].
  
 \bibitem{lfv2} 
  J.~Hisano, T.~Moroi, K.~Tobe, M.~Yamaguchi and T.~Yanagida,
  Phys.\ Lett.\ B {\bf 357}, 579 (1995)
  [hep-ph/9501407].

\bibitem{focus_org}
  J.~L.~Feng, K.~T.~Matchev and T.~Moroi,
  Phys.\ Rev.\ Lett.\  {\bf 84}, 2322 (2000)
  [hep-ph/9908309];
  J.~L.~Feng, K.~T.~Matchev and T.~Moroi,
  Phys.\ Rev.\ D {\bf 61}, 075005 (2000)
  [hep-ph/9909334].

\bibitem{focus}
G.~L.~Kane and S.~F.~King,
  Phys.\ Lett.\ B {\bf 451}, 113 (1999)
  [hep-ph/9810374];
H.~Abe, T.~Kobayashi and Y.~Omura,
  Phys.\ Rev.\ D {\bf 76}, 015002 (2007)
  [hep-ph/0703044 [hep-ph]];
   S.~P.~Martin,
  Phys.\ Rev.\ D {\bf 75}, 115005 (2007)
  [hep-ph/0703097 [hep-ph]];
   D.~Horton and G.~G.~Ross,
  Nucl.\ Phys.\ B {\bf 830}, 221 (2010)
  [arXiv:0908.0857 [hep-ph]];
  J.~E.~Younkin and S.~P.~Martin,
  Phys.\ Rev.\ D {\bf 85}, 055028 (2012)
  [arXiv:1201.2989 [hep-ph]];
 I.~Gogoladze, F.~.Nasir and Q.~.Shafi,
  Int.\ J.\ Mod.\ Phys.\ A {\bf 28}, 1350046 (2013)
  [arXiv:1212.2593 [hep-ph]];
    A.~Kaminska, G.~G.~Ross and K.~Schmidt-Hoberg,
  JHEP {\bf 1311}, 209 (2013)
  [arXiv:1308.4168 [hep-ph]].

\bibitem{martin_gaugino}
  S.~P.~Martin,
  Phys.\ Rev.\ D {\bf 89}, 035011 (2014)
  [arXiv:1312.0582 [hep-ph]].


\bibitem{OYY}
  Y.~Okada, M.~Yamaguchi and T.~Yanagida,
  Prog.\ Theor.\ Phys.\  {\bf 85}, 1 (1991);
   Y.~Okada, M.~Yamaguchi and T.~Yanagida,
  Phys.\ Lett.\ B {\bf 262}, 54 (1991);
    J.~R.~Ellis, G.~Ridolfi and F.~Zwirner,
  Phys.\ Lett.\ B {\bf 257}, 83 (1991);
  H.~E.~Haber and R.~Hempfling,
  Phys.\ Rev.\ Lett.\  {\bf 66}, 1815 (1991);
  J.~R.~Ellis, G.~Ridolfi and F.~Zwirner,
  Phys.\ Lett.\ B {\bf 262}, 477 (1991).

\bibitem{yy_focus1} 
  T.~T.~Yanagida and N.~Yokozaki,
  Phys.\ Lett.\ B {\bf 722}, 355 (2013)
  [arXiv:1301.1137 [hep-ph]].
  
  \bibitem{yy_focus2}
  T.~T.~Yanagida and N.~Yokozaki,
  JHEP {\bf 1311}, 020 (2013)
  [arXiv:1308.0536 [hep-ph]].
  
  \bibitem{ATLAS}
  The ATLAS Collaboration, ATLAS-CONF-2013-014 (ATLAS NOTE).
  
\bibitem{CMS}
The CMS Collaboration, CMS PAS HIG-13-005.

  
  \bibitem{PGU}
  T.~Yanagida,
  Phys.\ Lett.\ B {\bf 344}, 211 (1995)
  [hep-ph/9409329];
    T.~Hotta, K.~I.~Izawa and T.~Yanagida,
  Phys.\ Rev.\ D {\bf 53}, 3913 (1996)
  [hep-ph/9509201];
  T.~Hotta, K.~I.~Izawa and T.~Yanagida,
  Phys.\ Rev.\ D {\bf 54}, 6970 (1996)
  [hep-ph/9602439];
   J.~Hisano and T.~Yanagida,
  Mod.\ Phys.\ Lett.\ A {\bf 10}, 3097 (1995)
  [hep-ph/9510277];
    K.~I.~Izawa and T.~Yanagida,
  Prog.\ Theor.\ Phys.\  {\bf 97}, 913 (1997)
  [hep-ph/9703350];
  T.~Watari and T.~Yanagida,
  hep-ph/0208107;
   T.~Watari and T.~Yanagida,
  Phys.\ Rev.\ D {\bf 70}, 036009 (2004)
  [hep-ph/0402160].

\bibitem{pdg}
J. Beringer et al. (Particle Data Group), Phys. Rev. D86, 010001 (2012). 

\bibitem{ft_measure}
J.~R.~Ellis, K.~Enqvist, D.~V.~Nanopoulos and F.~Zwirner,
  Mod.\ Phys.\ Lett.\ A {\bf 1}, 57 (1986);
   R.~Barbieri and G.~F.~Giudice,
  Nucl.\ Phys.\ B {\bf 306}, 63 (1988).

\bibitem{feynhiggs}
    S.~Heinemeyer, W.~Hollik and G.~Weiglein,
  Comput.\ Phys.\ Commun.\ \ {\bf 124}, 76  (2000)
  [hep-ph/9812320];
  S.~Heinemeyer, W.~Hollik and G.~Weiglein,
  Eur.\ Phys.\ J.\ C\ {\bf 9}, 343  (1999)
  [hep-ph/9812472];
  G.~Degrassi, S.~Heinemeyer, W.~Hollik, P.~Slavich and G.~Weiglein,
  Eur.\ Phys.\ J.\ C\ {\bf 28}, 133  (2003)
  [hep-ph/0212020];
  M.~Frank, T.~Hahn, S.~Heinemeyer, W.~Hollik, H.~Rzehak and G.~Weiglein,
  JHEP\ {\bf 0702}, 047  (2007)
  [hep-ph/0611326].

\bibitem{feynhiggs_higher}
  T.~Hahn, S.~Heinemeyer, W.~Hollik, H.~Rzehak and G.~Weiglein,
  arXiv:1312.4937 [hep-ph].
  
\bibitem{suspect}
    A.~Djouadi, J.~-L.~Kneur and G.~Moultaka,
  Comput.\ Phys.\ Commun.\  {\bf 176}, 426 (2007)
  [hep-ph/0211331].
  
\bibitem{znr_nonanomalous}
  M.~Asano, T.~Moroi, R.~Sato and T.~T.~Yanagida,
  Phys.\ Lett.\ B {\bf 705} (2011) 337
  [arXiv:1108.2402 [hep-ph]].
  
\bibitem{vector_gaugino}
T.~Moroi, T.~T.~Yanagida and N.~Yokozaki,
  Phys.\ Lett.\ B {\bf 719}, 148 (2013)
  [arXiv:1211.4676 [hep-ph]].


\bibitem{atlas_3000}
The ATLAS Collaboration, ATLAS-PHYS-PUB-2013-002 (ATLAS NOTE).


\bibitem{martin_rge}
  S.~P.~Martin and M.~T.~Vaughn,
  Phys.\ Rev.\ D {\bf 50}, 2282 (1994)
  [Erratum-ibid.\ D {\bf 78}, 039903 (2008)]
  [hep-ph/9311340].











\end{thebibliography}
\end{document}